# On the interpretability and computational reliability of frequency-domain Granger causality

Luca Faes[a], Sebastiano Stramaglia[b], Daniele Marinazzo[c]

[a] *BIOtech, Dept. of Industrial Engineering, University of Trento, and IRCS-PAT FBK, 38010 Trento, Italy*
[b] *Dipartimento di Fisica, Università degli Studi Aldo Moro, Bari, and INFN, Sezione di Bari, 70123 Bari, Italy*
[c] *Department of Data Analysis, Ghent University, Ghent, B9000, Belgium*
E-mail: daniele.marinazzo@ugent.be

Stokes and Purdon [1] perform a critical evaluation of Granger-Geweke causality (GGC), a popular tool to assess directed interactions from multivariate time series [2]. Using simulations, they evidence computational and interpretational problems in the frequency-domain formulation of GGC, concluding that the notion of causality underlying GGC may yield misleading results, incompatible with the objectives of many neuroscience studies.

We support the message that GGC and lag-based data-driven methods in general cannot measure "causality" as intended elsewhere (see [2,3] for a thoughtful distinction). On the other hand, we think that these methods are dismissed in [1] based on a suboptimal (albeit often used) formulation, and we show that spectral GGC estimates can be obtained with a high computational reliability if proper estimation approaches are employed, and the interpretation of frequency domain causality measures is meaningful if spectral and causal information are properly combined[1].

The first simulation in [1] shows that spectral GGC cannot be reliably estimated even for simple systems, due to the modeling requirements. Repeating this simulation with the same parameters and data length, we confirm that the standard method of fitting separate full and reduced vector autoregressive (VAR) models returns spectral GGC estimates which display a strong bias (Fig. 1a) or a very large variability (Fig. 1b) depending on the choice of the model order. As explained in [1], this tradeoff between bias and variance arises from the incorrect representation of the reduced model as a VAR process of finite order. However, the problem can be overcome employing a state-space (SS) approach [4], allowing GGC computation in closed form from the SS parameters of the VAR process. This approach yields accurate spectral GGC estimates, closely following the expected profiles over the coupled directions, with negligible magnitude over the uncoupled direction (Fig. 1c).

The second simulation in [1] shows that, due to the independence of GGC from the intrinsic dynamics of the "receiver" process, the spectral GGC profiles linking this process to its putatively causal "transmitter" process are often misleading because different systems can have identical causality functions but different receiver dynamics. In Fig. 2 we confirm this result using the directed coherence (DC), a long-known spectral dependency measure [5] that for pairwise processes is analytically related to the spectral GGC [6]. However, this invariance property follows from the clear-cut interpretation of DC as the relative amount of spectral power that, at each frequency, arrives to the receiver from the transmitter [7]. We remark that the DC is also useful to fully recover the functional oscillatory structure of the observed processes, as it shapes the receiver spectrum revealing the portion of its spectral power that is "causally" due to the transmitter; this is depicted in the spectral decomposition of Fig. 2.

In conclusion, while thanking the Authors for pointing out some weaknesses of GGC measures, we think that proper formulations can provide meaningful results of directed dynamical influence, whose interpretation as "causal" is still bound to the knowledge and good faith of those who write (and read) related scientific literature.

---

[1] The code for running our analyses is based on the Matlab© scripts published with refs. [4,6,7] and is provided as supplementary material to this letter, as well as available on github https://github.com/danielemarinazzo/GC_SS_PNAS

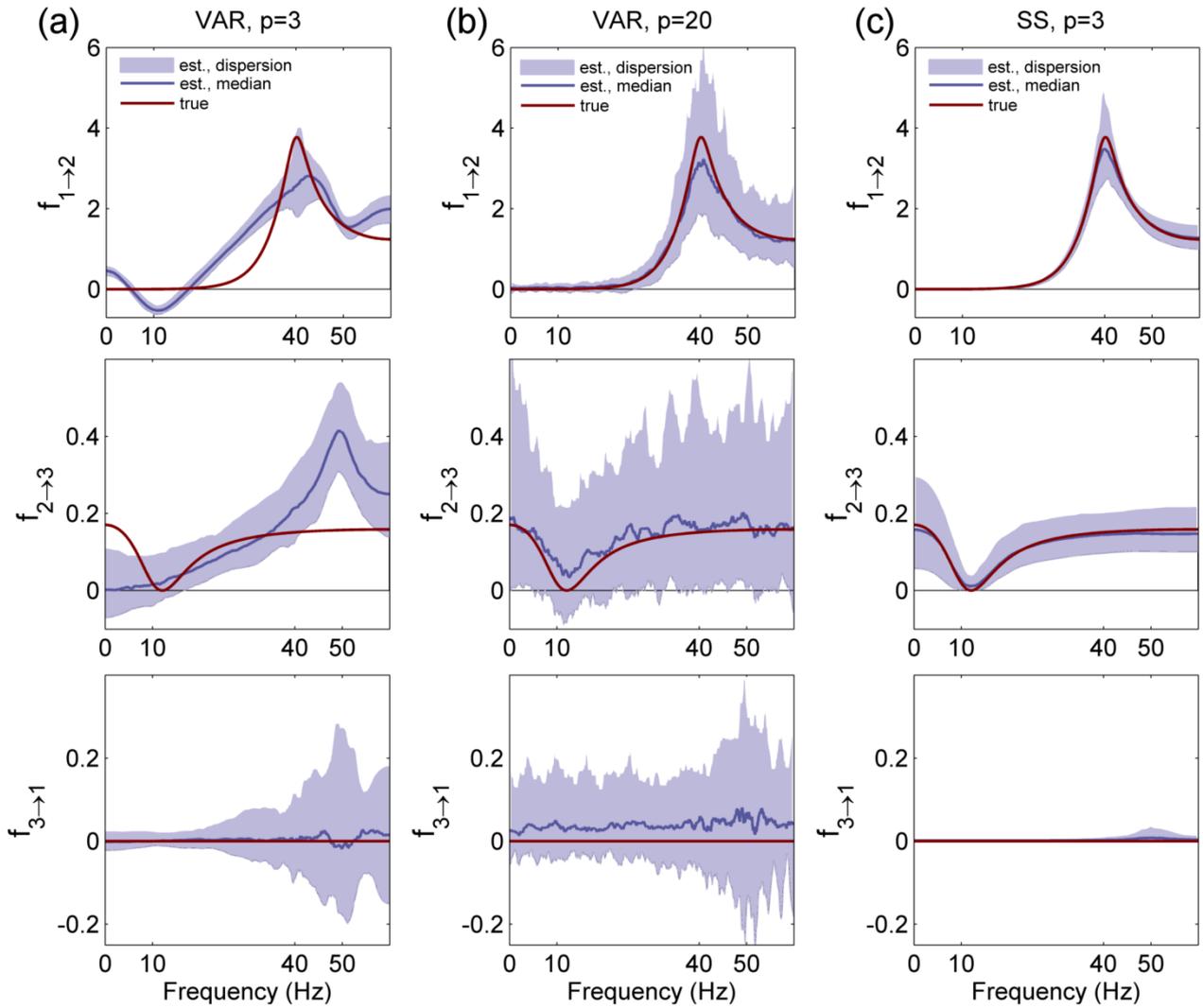

Fig. 1. Comparison of conditional frequency-domain Granger-Geweke causality (GGC) profiles computed for the three-node system of ref. 1 (Example 1, where nodes 1, 2, and 3 resonate respectively at 40 Hz, 10 Hz, and 50 Hz, and where unidirectional causality is imposed from node 1 to node 2, and from node 2 to node 3). GGC is computed along the two coupled directions ($f_{1\rightarrow 2}$, $f_{2\rightarrow 3}$) and along a direction with no coupling ($f_{3\rightarrow 1}$) using classical vector autoregressive (VAR) estimation of full and reduced models performed with the true model order p=3 (a) and with an increased order p=20 (b), and using state space (SS) estimation (c). The estimates reported as median and 5th-95th percentiles over 100 simulations (blue line and shades) show the lower variance of the SS method compared to the classical VAR approach. Moreover, the true causality values computed from the original model parameters (red lines) demonstrate the much lower bias of the SS estimates.

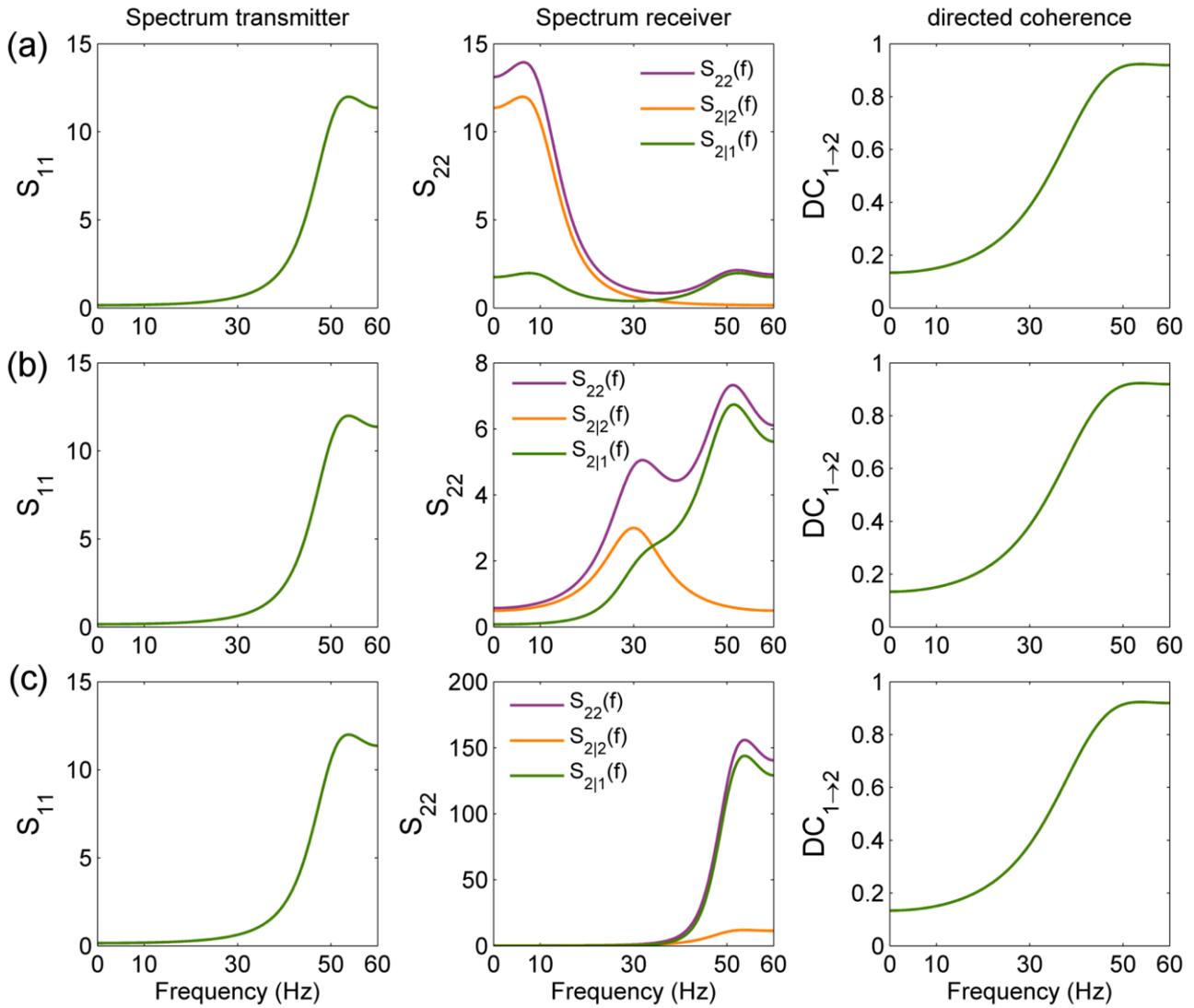

Fig. 2. Spectral and causality analysis of the two-node system of ref. 1 (Example 2, where unidirectional causality is imposed from node 1 to node 2) studied setting a resonance frequency of 50 Hz for the transmitter ($S_{11}$, green line in the left panels) and of 10 Hz (a), 30 Hz (b) and 50 Hz (c) for the receiver ($S_{22}$, purple line in the middle panels). The directed coherence $DC_{1\rightarrow 2}$ is the same for the three cases (green line in the right panels). However, it determines a different causal contribution of the transmitter on the receiver in terms of power spectral density $S_{2|1}=S_{22}\cdot DC_{1\rightarrow 2}$ (green line in the middle panels); the part not explained by the transmitter ($S_{2|2}$, orange line in the middle panels) reflects the autonomous dynamics of the receiver.